\documentclass[conference]{IEEEtran}
\IEEEoverridecommandlockouts
\usepackage{cite}
\usepackage{amsmath,amssymb,amsfonts}
\usepackage{algorithmic}
\usepackage{graphicx}
\usepackage{textcomp}
\usepackage{xcolor}
\def\BibTeX{{\rm B\kern-.05em{\sc i\kern-.025em b}\kern-.08em
    T\kern-.1667em\lower.7ex\hbox{E}\kern-.125emX}}
\begin{document}

\title{“If we didn't solve small data in the past, how can we solve Big Data today?”\\
{\footnotesize {Submission for CNIT 57000-001 DIS - Fall 2021}}

}

\author{\IEEEauthorblockN{Akash Ravi}
\IEEEauthorblockA{\textit{Dept of Computer and Information Technology} \\
\textit{Purdue University}\\
West Lafayette, USA \\
ravi48@purdue.edu}
}

\maketitle

\begin{abstract}

Data is a critical aspect of the world we live in. With systems producing and consuming vast amounts of data, it is essential for businesses to digitally transform and be equipped to derive the most value out of data. Data analytics techniques can be used to augment strategic decision-making. While this overall objective of data analytics remains fairly constant, the data itself can be available in numerous forms and can be categorized under various contexts. In this paper, we aim to research terms such as ‘small’ and ‘big’ data, understand their attributes, and look at ways in which they can add value. Specifically, the paper probes into the question “If we didn't solve small data in the past, how can we solve Big Data today?”. Based on the research, it can be inferred that, regardless of how small data might have been used, organizations can still leverage big data with the right technology and business vision. 

\end{abstract}


\section{Introduction}

In the 21st century, data assets are the driving engines behind the growth and value of organizations. Hence, it is critical to tap the potential of data to enhance the value provided to shareholders and drive innovation. Modern systems employ a host of techniques to perform descriptive, predictive, prescriptive, and diagnostic analytics on data. While the goals of each of these techniques focus on a different aspect, all of them rely on data. In most analytical and statistical methods, larger amounts of data equate to an accurate representation of the reality and improved generalizability of the results. However, in the real-life, collecting vast quantities of useful data might not always be easy. This brings us to the concept of big and small data. \\

\par
To start with an abstract understanding, big data is ‘bigger’ than small data in all measurable ways. While there are numerous terms to quantify big data, the volume, variety, and velocity of the data are the characteristic features used to draw the boundary between data and big data \cite{kitchin2016makes}. It is important to note that these boundaries are often fuzzy and relative. The categorization depends on the context, nature of the business/process, and the entities associated with the data. Big data often presents a host of concerns during processes such as data capture, storage, query, sharing, visualization, securing, and parallel processing. Due to the huge quantity, distributed computing methods are commonly used while handling big data to process them in a reasonable duration.  \\

\par
On the other hand, small data can be understood as quantities of data that can be comprehended by humans without sophisticated scientific tools and techniques. They focus on answering specific questions. Big data is predominantly focused on finding correlations, whereas small data is centered around finding the causation \cite{lindstrom2016small}. While large quantities of data are usually a prerequisite for data analytic methods to confidently uncover significant insights, small data can also be quite useful. It is easier to collect, mine, and make intuitive inferences. Small data has always existed. In the realm of computer science, they have been around since the times of punch cards \cite{bohme1991100}. Examples of small data include game scores, survey records, clinical measurements, and sales data. With the growth in computation power and storage options, capabilities around data management also grew. In the last decade, the development of cloud computing services and intelligent systems has fueled the concept of big data. Table \ref{big_vs_small_data} compares and contrasts the key features between big and small data based on how they differ relatively. While small data can also be high in volume or scope, big data usually check all the characteristics. \\

\begin{table}[]
\begin{tabular}{|p{0.25\linewidth}|p{0.3\linewidth}|p{0.3\linewidth}|}
\hline
\textbf{Features}     & \textbf{Big Data}             & \textbf{Small   Data}          \\ \hline
Volume                & Very Lage                     & Relatively   small             \\ \hline
Variety               & Wide                          & Narrow                         \\ \hline
Velocity              & Produced   continuously       & Non-continuous   collection    \\ \hline
Goal                  & Source for future   analytics & To answer   specific questions \\ \hline
Scope /   Resolution  & Fine-grained,   Exhaustive    & Focused                        \\ \hline
Relationality         & High                          & Depends on   context           \\ \hline
Ease of   Consumption & Difficult,   costly           & Easy, cheaper                  \\ \hline
\end{tabular}
\caption{Big Data vs Small Data}
\label{big_vs_small_data}
\end{table}

\par
Before we delve into the research question, it can be beneficial to understand examples of how organizations have dealt with big or small data in the past by surveying the available literature. This will provide insights into the potential competitive advantage that one could have when data is leveraged correctly. Further sections will also attempt to research how businesses can use prior knowledge to ‘solve’ big data today.

\section{Related Works}

Over the years there have been numerous studies on both big and small data. Researchers have studied and proposed novel ways to architect data pipelines and extract value from them. Different methodologies and algorithms have been developed to process various facets of data.  \\

\par
As it was previously mentioned, small data can also be extremely useful. Rob Kitchin and Tracey P. Lauriault \cite{kitchin2015small} have studied the ways in which the role of small data has been transforming with the development of big data. The authors present three key arguments. First, they show that studies related to small data will continue to hold their value and significance due to the proven record of the ability of small data to answer specific questions. Secondly, the developments happening around data processing for big data would also enhance how small data is handled. This would enrich the value of the data and hence, result in more useful structures being discovered. Since data-driven science uses a combination of abductive, inductive, and deductive methods to explain phenomena, modern analytic methods applied to traditional small data environments can help identify valuable insights that traditional scientific methods might have missed. Finally, the authors believe that scaling small data can help businesses uncover insights on interdisciplinary applications, that were not initially planned to be resolved. By combining small data with big data, we can increase the diversity of the data, thereby enabling us to engage in extensive and wide-ranging analytics. \\

\par
In a similar attempt to explain the potential benefits of small data, Julian et al. \cite{faraway2018small} researches on the trade-offs that would have to be made in real life while choosing big data over small data. The authors present multiple situations where small data could be used to arrive at an accurate conclusion in a faster, reliable, and cheaper way. First, the authors mathematically demonstrate the possibility of a smaller data-set being superior in quality compared to a larger one when the probabilistic sampling technique addressed bias sufficiently. The authors also show the ways in which the computational algorithms can be exponentially more expensive to perform as the size of the data continues to grow. In such cases, performing a comprehensive analysis on small data can lead to more desirable outcomes when compared against abstractly analyzing big data. The authors also point out that big data might pose major uncertainties, thus complicating statistical inferences. Concerns around factors such as data governance and privacy are also less pronounced in the case of small data. The authors also discuss the potential effects of aggregation bias. In some cases, the level of granularity provided by big data might not make sense in the real world. As long as the measurement intervals are sufficiently detailed, any further data would only be redundant. Finally, the authors discuss the advantages of small data with respect to teaching and instruction. Learning to statistically analyze and understand the business logic behind the data is easier with small data. \\

Huang et al. \cite{huang2015big} discuss another interesting observation. As previously seen, the distinction between big and small data is often fuzzy. The authors present the relationship between them. Big data can be decomposed into small data and small data can be aggregated into big data. There can be a mutual and co-dependent relationship between the two. The authors evaluate three models: one in which small data is a subset of big data, one in which they are independent, and the third model that is a hybrid between the two. Each of them has its own advantages and disadvantages. Small data can even be seen as the insight created from big data. They are hence difficult to evaluate as two separate topics. A lot of concepts indeed overlap between them.

\section{Was small data properly used?}

Despite seeing the numerous advantages that small data held, most organizations failed to leverage small data to its maximum extent. To have a quick recap of the history of computational data management, relational databases were first developed in the 1970s. Structured Query Languages (SQL) were developed to interact and transact with these databases. In the late 1980s, data storage started becoming cheaper and more accessible. This led to an increase in the volume of data be stored and processed. The concept of data warehousing was introduced to collate all the data originating from operational systems. Organizations started to host these data warehouses on mainframe systems. \\

\par
During the mid-1990s, the internet started growing in popularity and usage. Since relational databases were rigid in terms of structuring the data, there was an influx of unstructured data. NoSQL databases were developed for accommodating this need of having to store and use data in a flexible way. The term business intelligence (BI) has been around for a long time. However, data-driven decision-making was formalized and was computationally accessible only around this time. Another domain that grew with the growth of the internet includes data mining. It emphasized the process of extracting and discovering patterns from data sets. In 2007, it was found that BI was the top-most priority of many chief information officers \cite{watson2007current}. \\

\par
Despite the evolution of all these data processing and management techniques, most businesses were not digitized until the start of the 21st century. Even for those who had information systems during the early 2000s, the data being handled were mostly transactional. The adoption of data-driven decision-making increased multi-fold in the following decade \cite{brynjolfsson2016rapid}. Most organizations lacked the awareness of the potential of data prior to this period. The research on advanced data analytic techniques such as machine learning, deep learning, and artificial intelligence also increased significantly during this period \cite{batistivc2019history}. \\

\par
There are also a few inherent disadvantages with small data that did not let it succeed. Since small data was focused on answering specific questions, it did not provide enough scope for open-ended predictive and prescriptive analytics. The problem statements had to be pre-determined to a considerable extent before the data was collected. This made it difficult for organizations to venture into uncharted venues. Consequently, the impact created out of solving small data was usually non-disruptive and trivial. Due to the lack of a comprehensive end-to-end ecosystem for processing data, organizations had to deal with quality issues while trying to connect heterogeneous systems. Furthermore, since not all processes were digitized in the past, human error had a great effect on the resultant data quality \cite{redman1998impact}. Hence, managing and using data effectively posed significant challenges in the past.

\section{Big data challenges in today's world}

Before attempting to analyze the research question, it can be essential to understand the challenge posed by big data. This would help us comprehend what it takes to 'solve' big data. \\

\par
Sivarajah et al. \cite{sivarajah2017critical} groups the challenges in big data into three categories based on the data lifecycle: data, process, and management challenges. The first refers to the innate qualities of data such as volume, variety, and velocity. Process challenges span across data capture, integration, transformation, and algorithmic processing. Finally, management challenges deal with factors such as compliance, access control, privacy, infrastructure, and ethical concerns. Each of these challenges has a unique way of addressing them and has best practices associated with them. \\

\par
In addition to the large sample size, big data is also characterized by high dimensionality. This can cause issues such as noise accumulation, algorithmic instability, and heterogeneity \cite{fan2014challenges}. Modern systems produce petabytes of data within durations as short as a day. Scientific applications such as genomics, nuclear research, and modeling physical phenomena generate vast amounts of data and demand large amounts of computing power. Since the potential impact of these interdisciplinary research can affect our lives in unimagined ways, ethics is starting to be seen as a pillar that should be driving big data. Right from the data collection process to the decisions taken from the resultant analytics, the consequence of every action should be evaluated with empathy and a wider perspective. Subject matter expertise is required to oversee the big data projects and ensure that the goals are not overlooked when dealing with technology.\\ 

\par
As seen above, a few of the issues are due to technological constraints, while a few of them involve human decisions. As an example, the emergence of regulations such as the GDPR and the California Consumer Privacy Act has laid additional constraints on the ways in which data can be handled. Countries have also imposed data localization restrictions. While these regulations are absolutely necessary, they add additional steps in the process of working with big data. Such issues were much less pronounced back when small data dominated the industry. On the other hand, certain aspects have vastly improved. Cloud computing vendors provide various managed services to handle data across its lifecycle. A few tools do not even require the users to be proficient in mathematics or programming. Such a convenience was not available during the small data era. Most of the challenges in small data that were not related to the volume, velocity, or variety continue to exist even today. As a matter of fact, these issues have compounded due to the increasing complexity of big data. Thus, the deterrents have been evolving along with technology. Organizations would have to understand the unique contexts in which they use big data and mitigate against these hindrances to maximize the value. \\

\section{Revisiting the research question}

As it can be seen, despite all the advantages, small data did not create the impact that could have been achieved. Hence, we can state that a majority of organizations did not optimally 'solve' small data. In this context, solving abstractly refers to the process of addressing challenges and utilizing data to its maximum extent. Technologies around data analytics and data science have grown significantly. However, the challenges around 'solving' data have also transformed in various ways. \\

\par
On a positive note, data literacy has also increased over the years. Organizations have realized the importance of expediting digital transformation and leveraging big data to enhance business processes. Billions of dollars are being invested to support the data processing pipelines \cite{reed2015exascale}. Nevertheless, it is critical to understand the ways in which we are better equipped to 'solve' big data in today's world. The following subsections research the methods that could be instrumental in doing so.

\subsection{Technologies to solve big data}

Computational software to store, manipulate and analyze data have been around since the beginning of computers. However, we are now equipped with a set of robust and mature toolchains that can be adopted to architect our data processing solution. \\ 

\par
Apache Hadoop is a very famous open-source framework that provides highly scalable and reliable data processing services. The Hadoop Distributed File System (HDFS) delivers high throughput by storing and processing data in parallel environments. It can also integrate with various tools to ingest data and provide an interface for processing. Most big data tools rely on the concept of computing near the location of the data. Rather than moving data around, it is easier to compute the instructions on the nodes where the data resides. Concepts such as MapReduce is widely adopted to process data in this method. A MapReduce program is composed of a map phase, which involves filtering and sorting, and then a reduce phase, which executes a summary operation. These methods have made it extremely easy to perform operations on vast amounts of data. Apache Spark is another notable project that is often considered a successor to Hadoop. The in-memory computation capability in spark provides extremely fast big data processing. Other tools such as Kafka, Neo4j, Pentaho provides support for streaming data, visualization, etc. There are numerous open-source technologies that can be integrated to create a robust end-to-end data processing pipeline \cite{arfat2020big}.\\

\par
With the advent of cloud computing, there has been a paradigm shift in the ways big data can be solved. The biggest barrier to big data has been around technology. Most tools require specialized engineers and can involve significant research and development. However, cloud computing services have made it easier for companies to use these capabilities without having to deal with the nuances of the underlying technology. The level of abstraction can be customized based on the business needs. As an added advantage, most services contribute to operational expenses rather than an upfront capital expense. This can reduce the barrier inhibiting the process of solving big data \cite{yang2017big}. \\

\par
With so many tools and technologies available, there is no 'one-size-fits-all' tool-chain to create the perfect solution for solving big data. As an example, Brewer's CAP theorem poses an interesting concept to compare and contrast trade-offs that might have to be made in the process. The theorem states that a database system can simultaneously provide a maximum of two out of the following three guarantees: consistency, availability, and partition tolerance. Depending on the use case, the choice of the appropriate technology must be made. We would also have to determine the ways in which the system can scale. With most data pipelines being deployed on a cloud-native architecture, horizontal scalability has become essential to process vasts amounts of ever-growing data \cite{andrikopoulos2012designing}.

\subsection{Transforming business priorities}

While tools and technologies can simplify the process of 'solving' big data, their benefits provide tangible value only when it is understood and supported by the stakeholders involved. Similarly, big data can only provide actionable insights. Executing the needful is the responsibility of organizations. \\

\par
The Fourth Industrial Revolution (Industry 4.0) is the ongoing automation of manufacturing and industrial practices. Big data is a key aspect of enabling this transformation. Huge amounts of data are collected from smart sensors and IIoT platforms and stream to the cloud, where further analysis can be made to uncover patterns and improve efficiency. This can significantly reduce costs in the long run. A report from McKinsey claimed that leveraging big data in manufacturing can decrease product development costs by up to 50\% and reduce the working capital by 7\% \cite{yin2015big}. \\

\par
Even as more and more organizations are attempting to be data-driven, there can still be major obstacles that may impede progress towards data-oriented goals. A 2019 report from the Harvard Business Review found that a majority of organizations (greater than 50\%) agree that they are not competing on data and analytics and are yet to forge a culture that reflects on the importance of solving big data. The study also shows that only 7.5\% of business executives reported that technology was an obstacle. On the other hand, a whopping 93\% of the respondents agree that people and processes are the major factors that continue to pose a challenge in leveraging big data \cite{harvard_business_review_2019}. \\

\par
Business leaders need to recognize the need to invest in these technologies. Intellectual and capital expenditures made related to data should be viewed as investments rather than liabilities. Being proactive in identifying key metrics is essential to solving big data. The outcomes need to be aligned with what the business intends to accomplish. This vision would then have to be communicated across the organization to channelize their efforts accordingly. Finally, strategic decisions would have to be made based on the insights derived from the data. 

\section{Conclusion}

As we had seen, there are numerous applications and uses for data. They have the potential to transform industries, improve processes and aid the overall economic development in serendipitous ways. In the early 2000s, small data was the major type of data held by most organizations. Despite the potential of small data, it was not effectively 'solved' by many organizations. Firms that refused to embrace data and technology have suffered considerably. Without the competitive edge given data-driven decisions to steer the teams, few companies have even succumbed to the competition. \\

\par
As the sections elaborated, the concerns around leveraging data have also changed over the years. While a few issues have gotten worse, most of the advancements are geared towards enabling organizations to effectively solve big data. The current solutions are ought to get better by benefiting from economies of scale. Technologies are only going to get more powerful and as time progresses, the amount of accumulated data would also grow. Businesses must be able to understand the opportunities that have been missed in the past because of not solving small data effectively. These avenues should be re-evaluated to form a comprehensive big data strategy. Since it is better late than never, we could solve big data regardless of what the past has been. Big data continues to be researched and developed and hence, holds a lot of potential for the future. When approached methodologically, leveraging big data and embedding analytics into the core of business processes can help organizations scale and succeed. 

\bibliographystyle{ieeetr}
\bibliography{./ref}

\end{document}